\documentclass[aps,pre,twocolumn,groupedaddress,showpacs]{revtex4}
\usepackage{graphics,epsfig}
\usepackage{amssymb,amsfonts}
\usepackage{color}

\newcommand{\bin}[2]{{#1 \choose #2}}
\newcommand{\nn}{\nonumber}
\newcommand{\bbb}{\begin{eqnarray}}
\newcommand{\eee}{\end{eqnarray}}

\newcommand{\tx}{{\tilde x}}
\newcommand{\Gen}{g}
\newcommand{\Lap}{\hat f}
\newcommand{\re}{{\rm e}}
\newcommand{\gf}{{\cal F}}
\newcommand{\gff}{{\cal G}}
\newcommand{\betap}{\beta'}
\newcommand{\betaw}{\beta_{w}}
\newcommand{\aco}{a_{\infty}}

\begin{document}

\title{Weibull-type limiting distribution for replicative systems}

\author{Junghyo Jo$^1$, Jean-Yves Fortin$^2$ and M.Y. Choi$^3$}
\affiliation
{$^1$Laboratory of Biological Modeling, NIDDK, National Institutes of Health,
Bethesda, Maryland 20892, U.S.A.\\
$^2$Institut Jean Lamour, D\'epartement de Physique de la Mati\`ere et des Mat\'eriaux,
Groupe de Physique Statistique, CNRS - Nancy-Universit\'e
BP 70239 F-54506 Vandoeuvre les Nancy Cedex, France\\
$^3$Department of Physics and Astronomy and Center for Theoretical Physics,
Seoul National University, Seoul 151-747, Korea}

\begin{abstract}
The Weibull function is widely used to describe skew distributions
observed in nature. However, the origin of this ubiquity is not
always obvious to explain. In the present study, we consider the
well-known Galton-Watson branching process describing simple replicative
systems.
The shape of the resulting distribution, about which little has
been known, is found essentially indistinguishable from the
Weibull form in a wide range of the branching parameter; this can
be seen from the exact series expansion for the cumulative
distribution, which takes a universal form. We also find that the
branching process can be mapped into a process of aggregation of
clusters. In the branching and aggregation process, the number of
events considered for branching and aggregation grows cumulatively in
time, whereas for the binomial distribution, an independent event
occurs at each time with a given success probability.
\end{abstract}


\maketitle

\section{Introduction}
Various systems in nature exhibit skew distributions, which are properly fit to
the Weibull distribution \cite{Weibull} as well as lognormal and power-law distributions;
relations between those skew distributions have been discussed recently~\cite{Choi}.
In particular, the Weibull distribution,
despite the simple mathematical form, particularly for the
cumulative distribution $F(x) = 1-\exp[-(x/\eta)^\beta]$,
has flexible shapes depending on the value of $\beta$
and is widely used to describe size distributions of, e.g.,
material strengths~\cite{Weibull, Kirchner}, cloud droplets~\cite{Liu},
biological tissues~\cite{Jo}, ocean wave heights~\cite{Muraleedharan},
and wind speeds~\cite{Takle}.
However, there still lacks an appropriate explanation of its ubiquitous emergence,
in sharp contrast with the Gaussian distribution, let aside the case-by-case
derivation such as material breaking with the weakest element~\cite{Weibull},
entropy maximization~\cite{Liu}, material fragmentation~\cite{Brown},
and extreme value statistics~\cite{Bertin,Moloney}.

It is well known that the binomial distribution results from
success events for given independent trials with the success
probability $p$ given. When the success is a rare event (i.e., $p$
is small), it reduces to the Poisson distribution. According to
the central limit theorem~\cite{Kallenberg}, (discrete) binomial
and Poisson distributions approach the (continuous) Gaussian
distribution in the limit of large trial numbers. In a similar
spirit, we here derive a continuous Weibull-like distribution from
the discrete Galton-Watson branching process, motivated by cell
replication in a tissue~\cite{Jo}. The branching process can serve
as a basic model to describe discrete events having two
possibilities, e.g., replication/non-replication or
nucleation/non-nucleation. The generating function for this
distribution was first obtained in the seminal work of general
branching processes~\cite{harris48,seneta69}. Specifically,
asymptotics were derived in the more general case of multiple
replicates and extinction processes at each stage of the process,
added to possible immigration events (see for example Ref.
\cite{biggins93}), but little is known about the shape of the
distribution itself relatively to other standard distributions,
except for few very specific cases where the limiting distribution
can be computed exactly through the use of a rational form for the
generating function at the first stage of the process and which
usually leads to a simple exponential function. Here we find that
it is approached by the Weibull distribution in rather a wide and
realistic range of the replication parameter $p$, making the two
distributions surprisingly indistinguishable in practice.

This paper consists of four sections and an appendix.
In Sec. II, cell replication is described in terms of a branching process. 
The stationary distribution of the branching process is obtained and its general
properties are discussed. Results of Monte Carlo simulations are also presented.
Section III examines the relation between the distributions for different replication probabilities and probe the scaling with the help of an ansatz, which is justified from the exact series expansion. 
Finally, Sec. IV discusses and summarizes the results.
In Appendix, all the moments of the distribution are obtained analytically from
the recurrence relation of the generating function.

\section{Cell replication and branching process}

\begin{figure}
\centerline{\epsfig{file=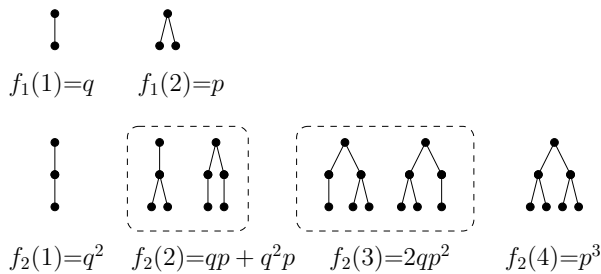,width=8.5cm}}
\caption{Cell replication graphs for a branching process. Cell number configurations at time steps $n = 1$ and $2$ are plotted with the replication probability at each step is given by $p$; $q \equiv 1-p$ corresponds to the probability that the cell does not replicate.}
\label{fig:branching}
\end{figure}

For the binomial distribution, an independent event occurs at each time with given success probability.
In cell replication, on the other hand, the number of replication events in consideration depends on the current cell number of a tissue.
For example, even if there exists just a single mother cell initially, it may replicate from time to time, and
there may occur many replications of the mother and daughter cells.
Accordingly, we consider the probability distribution $f_n(l)$ of tissues with size
(i.e., the number of cells) $l$ at given time step $n$, which
satisfies the normalization condition $\sum_{l=1}^{2^n} f_n(l) =1$ with the
maximum possible cell number in the tissue after the $n$th replication given
by $2^n$.
Note that this process can be described in terms of a branching process with the
branching probability $p$, as illustrated in Fig. \ref{fig:branching}.
Each graph in the figure, where sites in the $n$th row represent cells
at the time step $n$, corresponds to one possible configuration
of cell growth for the given duration.
Each graph thus starts from a single site in the first row (i.e., a single mother cell initially);
sites may replicate or not, giving birth to new sites
at successive time steps (here the time step is fixed to be a constant).

It is useful to consider the generating
function for the distribution $f_n$ at time $n$ in the branching
process \cite{harris48,Harris}:
\bbb
\Gen_n(z)=
\sum_{l=1}^{2^n}f_n(l)z^l.
\eee
For example, the $k$th moment
at time $n$, defined to be $\sum_{l=1}^{2^n}f_n(l)\,l^k$, can be
computed by differentiating successively the generating function:
$\left (z\frac{d}{dz}\right )^k\Gen_n(z)|_{z=1}$
with $\Gen_n(1) =1$ for all $n$
(see Appendix for the derivation of all the moments).
In the following, for simplicity, we will impose $f_n(l)=0$ for $l>2^n$.
At the initial time ($n=0$) the system contains only
one element, leading to $\Gen_0(z)=z$. Since the distribution $f_{n+1}$ is
related to the preceding one $f_n$ via combinatorial relations, it is easy to show
that the generating function satisfies the non-linear recursion equation,
$\Gen_{n}(z) = \Gen_1(\Gen_{n-1}(z))$
for $n \geq 1$, where $\Gen_1(z) = qz+pz^2$.
This equation provides a recursive function for the newly generated sites,
which are all independent,
with the generating function $\Gen_1 (z)$.

>From this relation, we can deduce that the total number $N(n)$ of configurations or graphs
at (discrete) time $n$ satisfies the recurrence relation $N(n{+}1)=N(n)[1+N(n)]$,
with the initial condition $N(0)=1$, and grows rapidly in time.
Indeed this relation can be obtained easily from the observation that
$N(n)$ is equal to $\Gen_n(1)$ with $p$ and $q$ replaced formally by unity.
Therefore $N(n)$ satisfies the same relation as $\Gen_n(1)$ above.
It is also manifested from the physical point of view:
Given $N(n)$ graphs at time $n$,
there are two possible ways to generate graphs at time $(n{+}1)$.
(i) In the case of non-replication of the original site,
we simply have $N(n)$ graphs; (ii) in the case of replication of the same site,
we can attach to the two offsprings a total of $N(n)^2$ graphs.
As a result, we obtain $N(n)+N(n)^2$ possible configurations at time $(n{+}1)$.
This can be checked in Fig.~\ref{fig:branching}
for the first few graphs: $N(0)=1$, $N(1)=2$, $N(2)=6$, and so on.
%
%

Because a tissue of size $l$ results from $(l{-}1)$-times proliferation
starting from a single cell (see Fig.~\ref{fig:branching}), the recurrence relation
\bbb
\Gen_{n+1}(z)=q \Gen_{n}(z)+ p \Gen_{n}^2(z)
\eee
leads to the recursive relation for the distribution $f_n(l)$
by simply identifying the coefficients of $z^l$ on the left and right sides of the last expression:
\bbb \label{fn}
f_{n+1}(l)=q f_n(l) + p \sum_{k=1}^{l-1} f_n(k) f_n(l-k).
\eee
Namely, a tissue of size $l$ at time $n+1$ can be generated
in the two ways:
(i) no replication at the first time step followed by producing $l$ descendants
at the following $n$ time steps
and (ii) replication at the first time step followed by
producing $k$ descendants from one offspring and
$l{-}k$ descendants from the other offspring at the following $n$ time steps.

\begin{figure}
\centerline{\epsfig{file=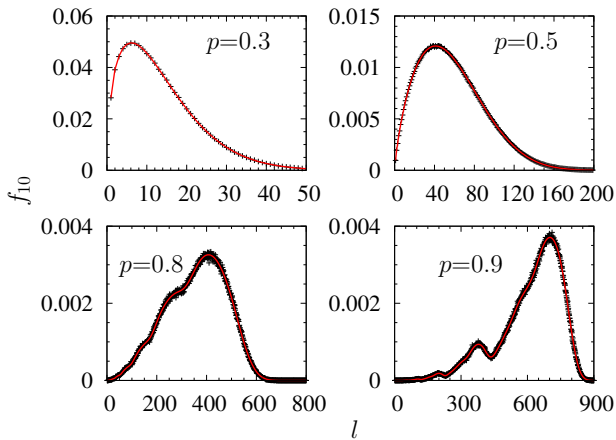,width=8.5cm}}
\caption{(color online) Comparison of the tissue size distribution $f_n (l)$ at time $n=10$,
for the replication probability $p = 0.3, 0.5, 0.8$, and $0.9$.
Analytical (solid lines) and simulation (+ signs) results agree perfectly,
displaying multimodal shapes for large values of $p$.
In the Monte Carlo simulations of the branching process,
starting from a single cell, we let every cell replicate with a given replication probability
at each Monte Carlo step. Data have been obtained from $10^6$ trial moves.
}
\label{fig:sizecomp}
\end{figure}

The size distribution, computed from Eq.~(\ref{fn}), is exhibited in Fig.
\ref{fig:sizecomp}, together with that from Monte Carlo
simulations, manifesting perfect agreement. It is of interest that
Eq.~(\ref{fn}) can be mapped into a process of random aggregation
of clusters with the aggregation probability $p$. Using $q=1-p$
and $\sum_{k=1}^{2^n} f_n(k) = 1$, we thus obtain
\bbb
\label{aggregation} \Delta f_n(l) = -p \sum_{k=1}^{2^n} f_n(l)
f_n(k) + p \sum_{k=1}^{l-1} f_n(k) f_n(l-k)
\eee
with $\Delta
f_n(l) \equiv f_{n+1}(l) -f_n(l)$. Therefore a cluster of size $l$
can be formed from aggregation of a cluster of size $k$ and a
cluster of size $(l-k)$ with the aggregation probability $p$.

\begin{figure}
\centerline{\epsfig{file=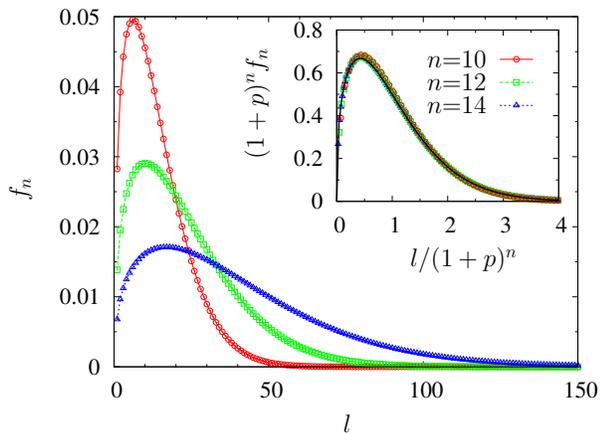,width=8.5cm}}
\caption{(color online) Weibull distribution of tissue sizes in the cell replication process.
Cell-number distribution for $p=0.3$ at three different time steps.
Distributions versus the rescaled size are plotted in the inset;
the collapse is fitted with a Weibull function with the shape parameter $\beta=1.37$ (black line).}
\label{fig:weibull}
\end{figure}

Figure \ref{fig:weibull} shows the normalized size distribution
for $p=0.3$ at several time steps $n=10, 12$, and $14$.
Remarkably, when size $l$ is rescaled by the factor $(1+p)^{n}$,
the distributions collapse into a single curve independent of $n$,
suggesting the presence of a stationary distribution for the
branching process~\cite{harris48}. Indeed the average cell number
in a tissue after the $n$th replication with the replication
probability $p$ is given by $(1+p)^n=\sum_{l =1}^{2^n} l f_n (l)$.
Note that $f_n(l)$ may be regarded as a continuous function
$f_n(x)$ when $n$ is large (see Fig. \ref{fig:weibull}).
Since the average cell number after the $(n{-}1)$th replication
is $(1+p)^{n-1}$, we have the scaling relation
\bbb
\int dx \,x f_{n}(x)&=&(1+p) \int dx'
\,x' f_{n-1}(x')
\nn \\
&=&\int dx\,(1+p)^{-1} x f_{n-1}((1{+}p)^{-1} x),
\eee
which is consistent with the fact that the distribution in the long-time limit
can be described by a time-independent stationary function 
$f(\tilde{x})$ with the rescaled size $\tilde{x} = x/\eta$
and the scale parameter $\eta= a (1+p)^n$.
The scale factor $a$ introduced here depends in particular on the replication
probability $p$ via boundary conditions, as discussed later.

Finally, a quantity of interest is given by the Laplace transform 
$\Lap(\lambda) \equiv \int_0^{\infty} d\tx \,e^{-\lambda \tx}
f(\tx)$, for which the recursive relation in Eq.~(\ref{fn}) reads
\cite{harris48} \bbb\label{laplac}
\Lap((1{+}p)\lambda)=q\Lap(\lambda)+p\Lap(\lambda)^2 . \eee
Equation (\ref{laplac}) takes the form of a Poincar\'e-type
equation \cite{lakner2008}, which is directly related in property
to Mahler functional equations \cite{mahler80} via an appropriate
change of variables \cite{poincare}.

In the limit of small $p$ where cells replicate very rarely, one
may expand Eq. (\ref{laplac}) as $\Lap((1{+}p)\lambda)\approx
\Lap(\lambda)+p\lambda \Lap'(\lambda)$, to obtain the differential
equation:
\bbb
\lambda
\Lap'(\lambda)=\Lap(\lambda)^2-\Lap(\lambda)
\eee
with the initial
conditions $\Lap(0)=1$ and $\Lap'(0)=-a^{-1}$. The solution reads
$\Lap(\lambda)=a(\lambda+a)^{-1}$, the inverse Laplace transform
of which is given by the simple exponential function
$f(\tx)=a\exp(-a\tx)$. With the constraint $F(1)=1-e^{-1}$ on the
cumulative distribution
$F(\tilde{x}) \equiv \int_0^{\tx} d\tilde{x}' f(\tilde{x}')$, we
obtain the scaling factor $a=1$ and therefore $f(\tx)=\exp(-\tx)$.
In the opposite case of $p=1$ where every cell replicates, we have
$\Lap(2\lambda)=\Lap(\lambda)^2$, with the simple solution
satisfying the initial conditions given by
$\Lap(\lambda)=\exp(-\lambda/a)$. This leads to the Dirac delta
distribution $f(\tx)=\delta(\tx-a^{-1})$ and the Heaviside
cumulative distribution $F(\tx)=\theta(\tx-a^{-1})$.
The constraint on $F(1)$ again imposes $a=1$.

\section{Scaling of the size distribution}
%
In this section, we consider the general case of $0<p<1$. As for the unique
stationary distribution $f(\tilde{x})$ for given $p$, one may
question whether there exists any relation between the
distribution $f(\tilde{x})$ corresponding to two different
replication probabilities $p$ and $p_0$, respectively. Since the
final stationary distributions result from the same branching
process, albeit with different branching probabilities, they are
expected to share qualitatively the same properties.

\begin{figure}
\centerline{\epsfig{file=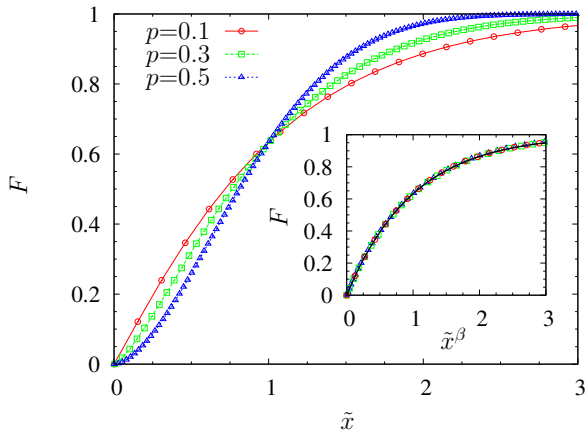,width=8.5cm}}
\caption{(color online)
Cumulative distribution for $p=0.1, 0.3$, and $0.5$.
The rescaled size is given by  $\tilde{x}= a^{-1}(1+p)^{-n} l$ with $n=20$, where
$a$ is the scale factor to adjust $F(1)=1-e^{-1}$.
Rescaled cumulative distribution functions are plotted in the inset,
disclosing the collapse into the function $F(\tx)=1-e^{-\tilde{x}^\beta}$ (black line).}
\label{fig:weibull2}
\end{figure}

To probe the scaling of the tissue size in the replication process,
we display in Fig. \ref{fig:weibull2} the cumulative distribution for
the replication probability $p=0.1, 0.3,$ and $0.5$.
Note that the scale factor $a$ in the rescaling of the size has been adjusted
to satisfy the condition $F(\tilde{x}{=}1)=1-e^{-1}$.
To probe the functional relations between the cumulative distributions 
for different values of $p$ 
under the constraints for $F$, we consider the change of variable
$\tilde{x}\rightarrow\tilde{x}^\beta$, as the simplest
possibility, where the exponent $\beta=\beta(p)$ is then adjusted
to make all curves for considered values of $p$ collapse onto a
single curve.
This ansatz indeed leads to the collapse of different cumulative
distributions into a unique distribution $F_0(\tilde{x})= 1 -
e^{-\tilde{x}}$, as shown in the inset. Therefore the new variable
$\tilde{x}^\beta$ determines the functional form of
$F(\tilde{x})$, at least for the numerical cases considered.
Indeed, using the known result $F(\tilde{x})=1-e^{-\tilde{x}}$ in
the limit $p\to 0$, we obtain
$F(\tilde{x})=1-e^{-\tilde{x}^\beta}$ with a good precision for
$p>0$, which leads to the Weibull distribution.

The ansatz of the scaling $\tx^\beta$ can be justified from the exact series expansion
of the distribution $f(\tx)$.
Multiplying both sides of Eq. (\ref{laplac}) by $\exp(-i\lambda
\tx)$, performing the rotation $\lambda \rightarrow i\lambda$, and
integrating over $\lambda$ along the real axis, we obtain
\bbb \label{func} 
\frac{1}{1+p}f((1{+}p)^{-1}\tx)=
qf(\tx)+p\int_0^{\tx}d\tx'f(\tx')f(\tx{-}\tx').
\eee
It can be
shown that $f(\tx)$ admits a series expansion in powers of $\tx$
consistent with the previous relation. In particular, $f(\tx)$
vanishes at the origin as $f(\tx) \approx a_0\tx^{\beta-1}$, with
some constant $a_0$ and exponent $\beta= - [\log(1{+}p)]^{-1}
\log(1{-}p)\ge 1$ \cite{harris48}. 
Here this analysis can be extended to consecutive terms to yield
the following expansion
\bbb\label{expf}
f(\tx)=\tx^{\beta-1}\sum_{k\ge 0} a_k \,\tx^{k\beta} ,
\eee
where
identifying the powers in Eq. (\ref{func}) gives the recursion
relation for the coefficients:
\bbb\label{ak}
(q^{k+1}- q) a_k = p\sum_{l=0}^{k-1}
B(\beta(1{+}l),\beta(k{-}l)) a_l \,a_{k-1-l}
\eee
with the beta function 
$B(x,y)=\int_0^1dt\,t^{x-1}(1-t)^{y-1}$.
Here $a_0$ is the only unknown parameter depending on boundary
conditions, since Eq. (\ref{ak}) implies the proportionality
relation $a_k\propto a_0^{1+k}$.

>From these results, it is easy to see that $f(\tx)$ can be cast
into the form
\bbb\label{scalingfunc}
f(\tx)=a_0\,\tx^{\beta-1}\gf(a_0\tx^{\beta})
\eee
with the
unique regular expansion of the scaling function:
$\gf(\tx)=\sum_{k\ge 0}\tilde a_k \tx^{k}$, where $\tilde a_k$
satisfies the relation in Eq. (\ref{ak}) but with the initial term
$\tilde a_0=1$; this determines uniquely all the other
coefficients $\tilde a_{k}$ for $k\ge 1$. The cumulative
distribution $F(\tx)$ is equal to a scaling function of the
variable $a_0\tx^{\beta}$ alone since
\bbb\label{cumulativeF}
F(\tx)
=\frac{1}{\beta}\sum_{k\ge 0}\frac{\tilde a_k}{k+1}(a_0\tx^{\beta})^{k+1}
=\gff(a_0\tx^{\beta}),
\eee
where $\gff$ is, like $\gf$, uniquely defined by the coefficients $\tilde a_k$.
The parameter $a_0$ is defined according to the constraint
$F(1)=1-e^{-1}$, and can be related to $a$ via the equation for
the first moment $\int_0^{\infty}d\tx \,\tx f(\tx) = a^{-1}$.
This relation simply gives
$a_0=a^{\beta}\left [\int_0^{\infty}u^{1/\beta}\gf(u)du \right]^{\beta}$.
Note that the cumulative distribution $F$ is a function of the
variable $\tx^\beta$ up to a scaling factor, which is also true for
the Weibull distribution, $F(\tx^\beta)=1-\exp(-\tx^\beta)$ with
$\tx=x/\eta$. In the limit of small $p$, $\beta$ is close to unity
and one can show that the expansion coefficients satisfying Eq.
(\ref{ak}) are approximatively given by $\tilde a_k=(-1)^k/k!$.
Therefore $\gff(a_0\tx^{\beta}) \approx 1-\exp(-a_0\tx^\beta)$ is
indeed close to the Weibull distribution.

The previous results show that the distribution can be expanded as
a series and vanishes as a power law with the exponent $\beta{-}1$
related to the replication probability $p$. In the opposite case
of large $\tx$, the integral equation (\ref{func}) can be
analyzed. Since we expect $f(\tx)$ to decrease with $\tx$ and
assume the stretched exponential behavior: $f(\tx)\approx
\exp(-\aco\tx^{\betap})$ with $\aco$ constant, we observe that in
Eq. (\ref{func}) the left-hand-side term $f((1{+}p)^{-1}
\tx)\propto \exp[-\aco (1+p)^{-\betap}\tx^{\betap}]$ is dominant
over the first term $f(\tx)$ on the right-hand side. The last term
can be analyzed by means of the saddle point analysis for the
function $\tx'^{\betap}+(\tx-\tx')^{\betap}$ appearing in the
exponential contribution. The saddle point, obtained by taking the
extremum of this quantity with respect to $\tx'$, corresponds to
the middle point of the integration $\tx'=\tx/2$. The overall
integral gives therefore a dominant contribution proportional to
$\exp[-2\aco(\tx/2)^{\betap}]$. The ansatz is consistent if the
two coefficients satisfy the relation
$(1+p)^{-\betap}=2^{1-\betap}$. This results in a new exponent
$\betap= \log 2 [\log 2 -\log(1{+}p)]^{-1}$ valid in the
asymptotic limit; this was also obtained in Ref. \cite{harris48}.

\section{Discussion}

\begin{figure}
\centerline{\epsfig{file=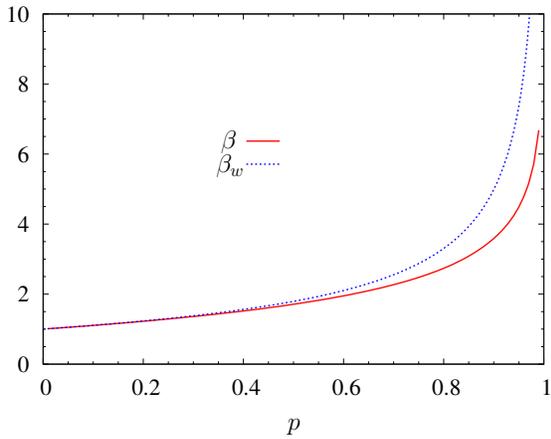,width=8.5cm}}
\caption{(color online) Relation between exponent $\betaw$ of the Weibull distribution and
the replication probability $p$ of the branching process.
The exponent $\beta$ is also plotted for comparison.}
\label{fig:exp}
\end{figure}

It has been shown that the replication process of cells with not
too large replication probability ($p \lesssim 0.5$) gives rise to
a distribution extremely close to the Weibull function. The
parameters of the Weibull distribution can then be related with
the first two moments of the distribution function $f_n(x)$:
$(1+p)^n = \eta \Gamma ( 1{+}\betaw^{-1})$ and $2(1+p)^{2n-1} =
\eta^2 \Gamma (1{+}2\betaw^{-1})$, where $\Gamma (x)$ is the Gamma function.
This leads to the following
relation between the replication probability $p$ and the shape
parameter $\betaw$ of the Weibull distribution:
\begin{equation}
p = 2 \frac{\Gamma^2(1{+}\betaw^{-1})}{\Gamma(1{+}2\betaw^{-1})} - 1 ,
\label{pbeta}
\end{equation}
which is exhibited in Fig. \ref{fig:exp}.
In addition, the scale factor $a$ in the rescaling parameter
$\eta=a(1+p)^n$ is given by $a=\Gamma^{-1}(1{+}\betaw^{-1})$. Note
that the exponents $\beta$ and $\betaw$ are hardly distinguishable
for $p \lesssim 0.5$, where the scaling function $\gf$ is
asymptotically similar to an exponential. This suggests that the
distribution in Eq. (\ref{scalingfunc}) belongs to the Weibull
class for small $p$. This regime applies to many cases in nature
that a certain event such as replication or nucleation occurs with
probability less than 50\,\% at a given time unit. On the other
hand, the replication process with a large value of $p$ results in
a different type of distribution, e.g., a multimodal distribution
(see Fig. \ref{fig:sizecomp}).

In conclusion, the branching process provides a general mechanism of the Weibull distribution
with $\beta \lesssim 2$, corresponding to the branching probability $p \lesssim 0.5$.
We have also found that the branching process can be mapped into a process of
aggregation of clusters.
A recent example includes the protein aggregation process with fission,
where the Weibull distribution with $\beta \sim 2$ emerges
as a stationary solution~\cite{Kunes}.

\begin{acknowledgments}
One of us (M.Y.C.) thanks Institut Jean Lamour, D\'epartement de Physique de la Mati\`ere et des Mat\'eriaux
at Universit\'e Henri Poincar\'e, where part of this work was carried out, for hospitality during his stay.
This work was supported by the intramural research program of NIH, NIDDK (J.J.)
and by NRF through the Basic Science Research Program (2009-0080791) (M.Y.C.).
\end{acknowledgments}

\appendix
*\section{Moment expression}

The functional equation, given by Eq. (\ref{laplac}), for the
Laplace transform of the size distribution can also be derived
with the help of moments of the distribution. Here we briefly
mention how to evaluate recursively all these moments starting
from the generating function.
>From the relation
\bbb
\langle x\rangle_{n}=z\frac{d}{dz}\Gen_{n}(z)|_{z=1}=zg_1'(\Gen_{n-1}(z))\Gen'_{n-1}(z)|_{z=1}
\eee
with the initial condition $\Gen'_0(1)=1$,
the average number of elements is simply:
\bbb\nn
\langle x\rangle_{n}=(1+p)\langle x\rangle_{n-1}=(1+p)^{n},
\eee
whereas the second moment is given by
\bbb
\langle x^2 \rangle_{n}=z\Gen_n'(z)+z^2\Gen''_n(z)|_{z=1}.
\eee
To evaluate $\Gen''_n(1)$, we differentiate the recursion relation for the generating function and obtain
\bbb\nn
\Gen_{n+1}''(z)=g_1''(\Gen_n(z))\Gen'^2_n(z)+g_1'(\Gen_n(z))\Gen''_n(z),
\eee
which leads to
\bbb\nn
\Gen_{n+1}''(1)=2p(1+p)^{2n}+(1+p)\Gen''_n(1).
\eee
Noting that
$\Gen_1''(1)=2p$ and $\Gen_0''(1)=0$,
%
we obtain the general solution of the previous recursion
\bbb\nn
\Gen''_n(1)=2\left[(1+p)^{2n-1}-(1+p)^{n-1} \right]
\eee
and the
second moment
\bbb\nn
\langle x^2 \rangle_{n} &=& 2(1+p)^{2n-1}-(1-p)(1+p)^{n-1} \nn \\
&\approx& 2(1+p)^{2n-1}.
\eee

In this large-$n$ (i.e., long-time) limit, one may define the scaling
relation
$\langle x^k\rangle_n  \simeq \Gen_n^{(k)}(1)\simeq h_k (1+p)^{kn}$,
where the first few coefficients read
\begin{eqnarray}
\label{firsth012}
h_0 = h_1=1,\;\;
h_2 = \frac{2}{1+p}.
\end{eqnarray}
For the $k$th moment $\langle x^k\rangle_n$,
given by a sum of derivatives of $\Gen_n$,
it is indeed sufficient to compute the largest (i.e., $k$th) derivative
of $\Gen_n$, which gives the essential contribution to the coefficient $h_k$.


%
A general method can be developed to evaluate the successive
moments by computing the dominant part of the derivatives of
$\Gen_n(z)$ in the large-$n$ limit. The $k^{th}$ derivative
$\Gen_{n}^{(k)}(z)$ satisfies indeed the following relation:
\bbb
\label{Gkn} \Gen_{n+1}^{(k)}(z) = g_1''(\Gen_n(z))\,T_{n,k}(z) +
g_1'(\Gen_n(z))\,\Gen_n^{(k)}(z)
\eee
with the initial conditions
$T_{n,1}(z)=0$, $T_{n,2}(z)=\Gen'^2_n(z)$, and
$T_{n,3}(z)=3\Gen'_n(z)\Gen''_n(z)$.
Taking the derivative of Eq. (\ref{Gkn}) with respect to $z$, we
obtain the relation for the coefficient $T_{n,k}(z)$:
\bbb
T_{n,k+1}(z)=\frac{\partial}{\partial
z}T_{n,k}(z)+\Gen'_n(z)\Gen_n^{(k)}(z).
\eee
This can be solved by
iterations
\bbb\nn
T_{n,k+1}(z)&=&\sum_{m=0}^{k-1}\frac{\partial^m}{\partial z^m}
\left[\Gen'_n(z)\,\Gen_n^{(k-m)}(z) \right]
\\ \label{Tnk}
&=&
\sum_{m=0}^{k-1}\sum_{l=0}^{m}\bin{m}{l}\,\Gen_n^{(l+1)}(z)\,\Gen_n^{(k-l)}(z),
\eee
where it has been noticed that
$T_{n,k}(z)$ contains at most the
$(k{-}1)$th derivative of $\Gen_n(z)$.

Since $g_1'(1)=1+p$ and $g_1''(1)=2p$, Eq. (\ref{Gkn}), together with Eq. (\ref{Tnk}),
bears the solution for $z=1$:
\begin{widetext}
\bbb \Gen_{n+1}^{(k)}(1) = 2p\sum_{j=0}^{n-1}(1+p)^j\,T_{n-j,k}(1)
               = 2p\sum_{m=0}^{k-2}\sum_{l=0}^{m}
                         \sum_{j=0}^{n-1}(1+p)^j\bin{m}{l}
                         \Gen_{n-j}^{(l+1)}(1)\,\Gen_{n-j}^{(k-1-l)}(1).
\eee
In the large-$n$ limit, we may use the scaling relation
$\Gen_n^{(k)}(1) = h_k(1+p)^{kn}$, so that the dependency on $n$ can be factorized,
which leads to the non-linear recursive relation for $h_k$:
\bbb\label{moments} h_k =\frac{\langle x^k \rangle_n}{(1+p)^{kn}}
= \frac{2p}{(1+p)^k-(1+p)}\sum_{m=0}^{k-2}\sum_{l=0} ^ { m }
\bin{m}{l} h_{l+1}h_{k-1-l}. \eee
This equation, together with the low-order coefficients in Eq. (\ref{firsth012}),
gives all the successive coefficients by simple iterations.
\end{widetext}

>From the nonlinear relations in Eq. (\ref{moments}), one can
reconstruct directly the Laplace transform of the stationary
distribution in Eq. (\ref{laplac}):
\bbb\label{Ll}
\Lap(\lambda)\equiv \int_0^{\infty}d\tx\,e^{-\lambda \tx} f(\tx)
=\sum_{k\ge 0}\frac{(-\lambda)^k}{a^kk!}h_k , \eee for which the
functional equation can be obtained.

In addition, Eq. (\ref{expf}) gives directly the exact large-$\lambda$ behavior of the Laplace transform $\Lap(\lambda)$ (see also \cite{harris48}),
which can be written as
\bbb\nn
\Lap(\lambda)&=&\int_0^{\infty} d\tx \re^{-\lambda \tx}f(\tx) = \sum_{k\ge 0}a_k\int_0^{\infty}\re^{-\lambda \tx}\tx^{\beta(k+1)-1}d\tx
\\ \label{Hl}
&=&\sum_{k\ge 0}a_k\frac{\Gamma(\beta(k{+}1))}{\lambda^{\beta(k+1)}}
\stackrel{\lambda\gg 1}{\approx}
\frac{a_0\Gamma(\beta)}{\lambda^{\beta}}.
\eee

\section*{References}
\bibliography{weibull}

\end{document}